\documentclass[preprint]{iucr}
\journalcode{S}

\usepackage[utf8]{inputenc}
\usepackage[margin=25mm]{geometry}
\usepackage{amsmath}
\usepackage{amsfonts}
\usepackage{amssymb}
\usepackage{mathtools}
\usepackage{graphicx}
\usepackage{xcolor}
\usepackage{url}
\usepackage{algorithm2e}

\DeclareMathOperator*{\argmin}{arg\,min}

\providecommand{\keywords}[1]
{
  \small	
  \textbf{\textit{Keywords---}} #1
}
\begin{document}

\title{tomoCAM: Fast Model-based Iterative Reconstruction via GPU Acceleration and Non-Uniform Fast Fourier Transforms}

\cauthor[a,c]{Dinesh}{Kumar}{dkumar@lbl.gov}{}
\author[b,c]{Dilworth Y.}{Parkinson}
\author[a,c]{Jeffrey J.}{Donatelli}

\aff[a]{Mathematics Department, Lawrence Berkeley National Laboratory, \city{Berkeley}, CA}
\aff[b]{Advanced Light Source, Lawrence Berkeley National Laboratory, \city{Berkeley}, CA}
\aff[c]{Center for Advanced Mathematics for Energy Research Applications, Lawrence Berkeley National Laboratory, \city{Berkeley}, CA}
\date{} 

\maketitle
\begin{abstract}

    X-Ray based computed tomography (CT) is a well-established technique for 
    determining the three-dimensional structure of an object from its two-dimensional 
    projections. In the past few decades, there have been significant advancements in the brightness 
    and detector technology of tomography instruments at synchrotron sources. These advancements have led to 
    the emergence of new observations and discoveries, with improved capabilities such as faster frame rates, 
    larger fields of view, higher resolution, and higher dimensionality. These advancements have enabled the material science community to expand the scope of tomographic measurements towards increasingly \emph{in-situ} and \emph{in-operando} measurements.
    In these new experiments, samples can be rapidly evolving, have complex geometries, and restrictions on the field of view, 
    limiting the number of projections that can be collected.
    In such cases, standard filtered back-projections (FBP) for the reconstructions often result in poor-quality reconstructions. 
    Iterative reconstruction algorithms, such as model-based iterative reconstructions (MBIR),
    have demonstrated considerable success in producing high-quality reconstructions under such restrictions, but typically require high-performance computing resources with hundreds of
    compute nodes to solve the problem in a reasonable time. 
    
   
    Here, we introduce
    \textbf{tomoCAM}, a new GPU-accelerated implementation of model-based iterative reconstruction, that leverages non-uniform
    fast Fourier transforms (NUFFTs) to efficiently compute Radon and back-projection operators and asynchronous memory transfers to maximize the throughput to the GPU memory. The resulting code is significantly faster than traditional MBIR codes and delivers the 
    reconstructive improvement offered by MBIR with affordable computing time and
    resources. \textbf{tomoCAM} has a Python front-end, allowing access from Jupyter-based
    frameworks, providing straightforward integration
    into existing workflows at synchrotron facilities.
    
\end{abstract}

\section{INTRODUCTION}
Micro- and nano-tomography using synchrotron technology is crucial in uncovering the inner makeup of modern materials, particularly in dynamic settings. Its diverse applications include the investigation of the fractures and deterioration of ceramic matrix composites, which are novel lightweight materials used in jet engines that operate under high temperatures and pressure \cite{Forna-Kreutzer2021}; the study of the flow of oil, brine, and carbon dioxide through rocks \cite{Walsh2014}; and the analysis of dendrite formation in batteries, which causes capacity reduction and eventual failure \cite{Lara2023}. Many synchrotron micro-CT facilities now have 
cameras that can acquire many-megapixel images at thousands of 
frames per second \cite{BL_ALS832, BL_APS2BM, BL_NSLS2TXM, BL_PSITOMCAT, Mokso2017}. 
These advances in instrumentation have encouraged users to push the 
boundaries of  what can be imaged at synchrotron beamlines. An increasing number of 
investigators are conducting \emph{in-situ} \cite{Larson2018, French2022} and \emph{in-operando} \cite{Kulkarni2020, Lara2023} measurements.
Typically the initial technique attempted for micro-CT reconstructions is the 
\emph{filtered back-projection} (FBP) method \cite{Herman2009}, which is available in \emph{tomopy} \cite{tomopy}  and \emph{tomocupy} \cite{Nikitin2023}. However, in the case of many dynamic experiments, where the specimen under observation is changing rapidly, 
it is generally not possible to capture sufficient projections to satisfy angular Shannon sampling conditions \cite{Crowther1970}
and overcome the noise. 
FBP is not a suitable option in such situations. The reconstructions obtained through this method tend to have excessive 
noise levels and exhibit streaking artifacts, 
making it difficult or even impossible to carry out further analysis.

As an alternative, iterative methods, such as
simultaneous iterative reconstruction technique (SIRT) \cite{Tarantola1982} and 
model-based iterative reconstruction (MBIR) \cite{Venkat2013, Aditya2014} 
aim to mitigate these shortcomings. 
They formulate the reconstruction as an optimization problem. The solution is obtained through an iterative process that aims to minimize the mismatch between the measured data and a forward model (Radon transform) of a digital representation of the sample. This iterative approach enables the incorporation of prior knowledge, such as total-variation constraints, into the optimization process, as demonstrated in various studies \cite{Trampert1990, Zhang2014, Venkat2013, Aditya2014}. However, current CPU-based implementations of MBIR typically require a large compute cluster to achieve turnaround times that are comparable to data collection times. This not only adds extra time to the experiment-to-analysis loop but also places an additional burden on material scientists, who must acquire a new set of expertise in using a compute cluster. This paper introduces \textbf{tomoCAM}, a GPU-accelerated implementation of MBIR that is based on the NUFFT approach. With the computational power provided by modern GPU devices and the relatively affordable cost of computer memory, it has become possible to perform these reconstructions on a single machine within a reasonable amount of time.

To design our GPU-accelerated algorithm and implementation, we build Radon and back-projection 
operators based on Non-uniform Fast Fourier Transforms (NUFFT) \cite{Greengard2004, Fessler2003},
which significantly reduces the computational cost, while maintaining high accuracy. 
We also leverage highly optimized \texttt{cuFFT} libraries that are native 
to the \textbf{CUDA} software development kit \cite{cuda}. We follow the mathematical outline 
laid out in \cite{Venkat2013, Aditya2014} to add a
\emph{total variation} constraint, which helps in reducing noise while preserving the sharp edges.
An important feature of our implementation is the flexibility to introduce a different constraint.
The choice of the constraint is not limited by
the algorithm design.

We test our computational framework through a series of numerical experiments on known phantoms and experimental data made publicly available through 
 Tomobank \cite{tomobank2018}. We compare the reconstructions with those 
obtained from  filtered back-projection \cite{tomopy} and 
\textbf{SVMBIR} \cite{svmbir2020}, a publicly available CPU-based package. 

In our numerical experiments, we found that:
\begin{itemize}
    \item when compared to FBP, the reconstruction quality produced by \textbf{tomoCAM} was superior with less noise, and required a lower number of projections,
    \item additionally, we observed that \textbf{tomoCAM} was around 15 times faster on a single machine than \textbf{SVMBIR}.
\end{itemize}. 

Finally, we provide a python front-end that exposes \textbf{tomoCAM}
functionality to the widely-used Numpy package \cite{numpy}, 
using \texttt{Pybind11} \cite{pybind11}. This makes it easy 
to integrate tomoCAM into existing workflows. The code is freely available at
\texttt{https://github.com/lbl-camera/tomocam}.

\section{Radon Transform and Non-uniform Fast Fourier Transform}
This section provides a brief overview of the fundamental concepts related to tomography, including the Radon transform (as well as its adjoint) and its connection to the Fourier transform. To perform tomographic measurements, a series of images, referred to as projections, are captured at various angles by rotating either the camera or the sample being studied.

\subsection{Radon Transform}
\begin{figure}
\centering
\includegraphics[width=0.25\linewidth]{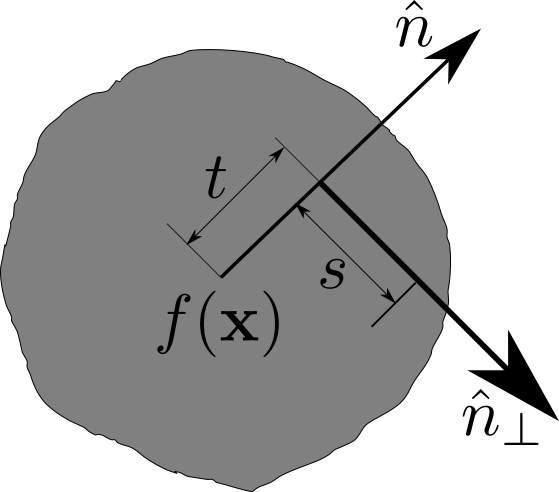}%
\caption{Radon transform of $f$ is its line integral along each line perpendicular to $\hat{n}$}
\label{fig:radon_transform}
\end{figure}

The Radon transform is fundamental to any tomographic reconstruction. It transforms a 
function $f(\mathbf{x},z),~ \mathbf{x} \in \mathbb{R}^2, ~z \in \mathbb R$, to $R\,f(t, \hat{n}, z), ~t \in \mathbb{R}, ~\hat{n} \in \mathbb{S}^1$ 
through a line integral \eqref{eqn:radon_transform}, see fig. \ref{fig:radon_transform}.
Given a set of oriented lines $\ell_{t, \hat{n}}$ defined as
\begin{equation}
    \ell_{t, \hat{n}} = \{ \mathbf{x} : \langle \mathbf{x}, \hat{n} \rangle  = t \} = \{ t \hat{n} + s \hat{n}_\perp : s \in \mathbb{R} \} \label{eqn:line_eqn}
\end{equation}
where $\hat{n}_\perp$ is direction of the X-ray beam, $\hat{n}$ is perpendicular to beam in same plane as $\ell$, and $t$ is the distance to $\ell$ from the origin. The Radon transform  $R f$ of function $f$ is defined as,

\begin{equation}\label{eqn:radon_transform}
    R f(t, \hat{n}, z) = \int_{\ell_{t,\hat{n}}} f(\mathbf{x},z)  = \int_{-\infty}^{\infty} f(t \hat{n} + s \hat{n}_\perp, z)\,ds 
\end{equation}

Tomographic measurements can be accurately modeled as the Radon Transform of the sample density represented by $f$.
It is the inversion of equation \eqref{eqn:radon_transform} that reconstructs $f$ from the data,
and is of primary importance in tomographic reconstruction.
By the \emph{central slice theorem}, the Fourier transform of the Radon transform of $f$ in direction $\hat{n}$ is equivalent to the Fourier transform of $f$ along $\hat{n}$, i.e.,

\begin{equation} \label{eqn:fourier_radon}
\begin{aligned}
\mathbb{F}_1[R\,f](k, \hat{n}, z) &:= \int_{-\infty}^{\infty} e^{-2 \pi i k t} R\,f(t,\hat{n}, z) dt\\
&= [\mathbb{F}_2\,f](k \hat{n}, z),
\end{aligned}
\end{equation}
where $z$ is the dimension along the axis of rotation and $\mathbb{F}_d$ denotes the $d$-dimensional Fourier transform. Radon transform and its adjoint are two-dimensional operators that are applied slice-by-slice on three-dimensional data. For simplification, we will drop the $z$ dependency from the subsequent notations. Assuming $f$ and $\mathbb{F}_1\,f$ are integrable everywhere,
 the inverse of the Radon transform \eqref{eqn:radon_transform} is given by,

\begin{equation} \label{eqn:inverse_radon}
f(\mathbf{x}) = \int_0^{\pi} \int_{-\infty}^{\infty} 
e^{2\pi i k \langle \mathbf{x}, {\hat{n}(\theta)} \rangle} 
\int_{-\infty}^{\infty} e^{-2 \pi i k t} y(t, \hat{n}(\theta)) dt\, |k|\,dk\,d\theta,
\end{equation}
where we denote the $\theta$ dependency as $\hat{n}(\theta) = (\cos \theta, \sin \theta)$.
 
It is computationally very expensive to exactly compute equation \eqref{eqn:inverse_radon}. In practice, \eqref{eqn:inverse_radon} is efficiently approximated with a Non-Uniform Fast Fourier Transform \cite{tomopy} or directly estimated in real space through the use of various filters such as Shepp-Logan \cite{Shepp1974}, Ram-Lak \cite{RamLak1971}, and Butterworth \cite{Butterworth1930}, which approximate and weight by the Fourier sampling density, hence the name \emph{Filtered Back-projection}.
These filters are additionally designed to dampen out the higher Fourier frequencies.
This is the most commonly used method in the reconstruction of tomographic data, in part 
because of the sheer speed by which the inversion can be performed. However, in cases when the view is partially blocked, 
or the specimen is evolving, it may not be possible to collect enough projections to sufficiently sample the Fourier space.
In such cases, the Filtered Back-projection results in poor image quality.

\subsection{Non-Uniform Fast Fourier Transform (NUFFT)} \label{sec:nufft}
On a discrete uniform grid, inversion of the Radon transform entails computing Fourier coefficients along radial lines using a 1-dimensional Fast Fourier Transform (FFT), followed by 2-dimensional 
backward Fourier transforms from a non-uniform polar grid $\{\mathbf{k}_j = k_j (\cos\theta_j, \sin\theta_j) \}$ onto a Cartesian grid $\{(x_n,y_n)\}$, which can be represented as the summation

\begin{equation} \label{eqn:fourier_sum}
f_n = \sum_{j=1}^M c_j e^{2 \pi i k_j (x_n\,\cos\theta_j + y_n\,\sin\theta_j)}, ~~~~ n \in [1, N],
\end{equation}
where $c_j$ is the Fourier coefficient at $\mathbf{k}_j$, $N$ is number of discrete points that represent the sample density $f$ on a uniform Cartesian grid,
and $M$ is the number of polar grid points representing the projection data.
However, directly computing equation \eqref{eqn:fourier_sum} is computationally expensive, as the complexity is $\mathcal{O}(M N)$.
Data taken at synchrotron light-sources can usually reach up to $M = \mathcal{O}(10^{10})$ pixels, and the final reconstructed image size $N$ has a similar 
order or magnitude for the final reconstructed image.

Non-Uniform Fast Fourier transforms (NUFFTs) offer a precise and efficient method for computing equation \eqref{eqn:fourier_sum}. This method involves first computing the Fourier coefficients on a polar grid using a sequence of one-dimensional FFTs along radial lines. Then, the computed coefficients are convolved with a compactly supported spreading kernel $\varphi$, and this convolution is evaluated on a uniform grid. Subsequently, an inverse Fourier transformation is performed on the convolution values on the uniform grid, followed by division by the Fourier transform $\hat{\varphi}$ of the kernel, i.e.,
\begin{align} \label{eqn:nufft}
    c_j =& \sum_t \rho(t, \theta_j) e^{-2 \pi i t k_j} \\ 
    F_r =& \smashoperator{\sum_{\|\mathbf{k}_r^c - \mathbf{k}_j\| < W}} c_j \varphi(\mathbf{k}_r^c- \mathbf{k}_j),\label{eqn:nufft2} \\
    f_n \approx& ~ \hat{\varphi}^{-1} {\mathbb{F}_2}^{-1}(F)\label{eqn:nufft3}
\end{align}
where $\{\mathbf{k}_r^c\}$ is a Cartesian grid, $\mathbb F_2$ is the 2D Fourier transform, equations \eqref{eqn:nufft} and \eqref{eqn:nufft3} are computed via fast Fourier transforms, and $W$ is the spreading width of the convolution in equation \eqref{eqn:nufft2}. For an appropriately chosen kernel, the NUFFT has an error of $\epsilon$ if $W$ is chosen to span approximately $w = \log_{10}(1/\epsilon)$ grid points per dimension. The computational complexity of equations \eqref{eqn:nufft}-\eqref{eqn:nufft3} is $\mathcal{O}(M log M_t + w^2 N + N log N)$, where $M_t$ is the number of points in the radial direction of the polar grid. Since $w^2 \ll M$, this results in a massive speedup compared to the direct computation of equation \eqref{eqn:fourier_sum}.
The Radon transform can similarly be computed by performing the above steps in reverse order.
In this work we have used the \texttt{cuFINUFFT} \cite{Shih21} library to compute NUFFTs. 
For a detailed discussion on the topic, we refer the reader to \cite{Dutt1993FFT, Fessler2003, Greengard2004, Barnett19, Barnett21}. 

\begin{figure}
\centering
\includegraphics[width=0.75\textwidth]{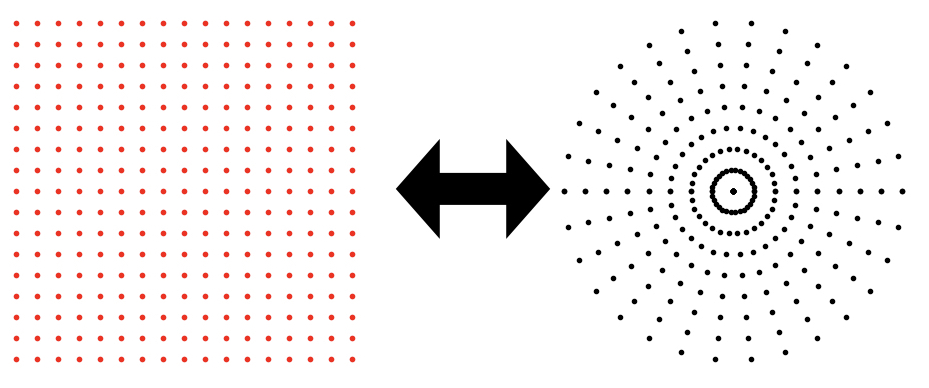}
\caption{The NUFFT is used to transform intensity on a uniform grid to its Fourier transform on a polar grid and \emph{vice-versa}.}
\label{fig:gridding}
\end{figure}

\section{Model-based Iterative reconstruction}

An alternate approach to FBP methods is to rely on iterative methods, such as MBIR.
Although these methods have longer processing times, they produce better-quality reconstructions
when compared with FBP methods. This is especially noticeable when a 
smaller number of projections are available.  This is because iterative methods are 
able incorporate \emph{a priori} information as a constraint on the optimization process.
We refer the reader to \cite{ASTME1441} for a more detailed discussion.
Iterative methods seek a solution $f$ by minimizing the difference between its Radon transform and projection data $b$, i.e., 
\begin{equation}
    f = \argmin_{f}\, \underbrace{\| R f - b \|^2}_{\mathcal{G}} 
    + \underbrace{ g(f)}_{\mathcal{H}}.
    \label{eqn:objfn}
\end{equation}
 
Here we set up the objective function as a least-squared problem.
The target is to iteratively search for $f$ that minimizes the $\ell^2$-norm of difference between $Rf$ and $b$ while penalizing violation of the constraint by $g$.
Now we differentiate equation \eqref{eqn:objfn} with respect to $f$ and equate the result to 0.
The gradient of the first term is 
 \begin{equation} \label{eqn:gradient}
 \nabla \mathcal{G} =  R^* \left( R\,f - b \right),
  \end{equation}
where $R^*$ is the adjoint of equation \eqref{eqn:radon_transform},
 \begin{equation} \label{eqn:transpose}
 R^*\rho(\mathbf{x}) = \int_0^{\pi} \int_{-\infty}^{\infty} e^{2\pi i k \langle \mathbf{x}, \hat{n}(\theta) \rangle} 
\int_{-\infty}^{\infty} e^{-2 \pi i t k}\rho(t, \hat{n}(\theta)) dt\, \,dk \,d\theta,
 \end{equation}
which is simply equation \eqref{eqn:inverse_radon} without the scaling $|k|$. The operators $R$ and $R^*$ can be efficiently computed using NUFFT.

 In the results presented here, we choose $\mathcal{H}$ to be a total-variation penalty in equation \eqref{eqn:objfn}.  We follow the mathematical approach presented in \cite{Venkat2013, Aditya2014} to model $\mathcal{H}$ as a q-Generalized Gaussian Random Field (qGGRMF),

\begin{align}
    g_m &= \sum_{n} w_{mn} h_{mn}, ~~~  \forall  n \in \{ n ~|~ \| m - n \|_\infty \le 1 \} \label{eqn:reg2} \\
    h_{mn} &= \frac{\left(\frac{\left|f_m - f_n \right|}{\sigma}\right)^2}{c + \left(\frac{ \left|f_m - f_n \right|}{\sigma}\right)^{2-p}} \label{eqn:reg3}
\end{align}
where $h$ is defined over 1-hop neighborhood of $m$, with  $m$ and $n$ being integer coordinates on the three-dimensional uniform grid.
The weights $w_{mn}$ are the Gaussian weights that partition the unity and are inversely proportional to the distance
between $m$ and $n$. Hyper-parameters $c$, $p$, and $\sigma$ are used to control the strength of the penalty term. The term $\mathcal{H}$ is an algebraic expression, and can easily be differentiated. 

In this work, we have used a monotonic accelerated gradient method with restart detailed in \cite{GiselssonB14b},
 but it is possible to use other optimizers.

\section{Implementation}
 When it comes to implementing software solutions, performance is a critical factor. In this section, we discuss some 
 of the important implementation details that have a significant impact on the performance of \textbf{tomoCAM}. 
 These include factors such as memory management, and hiding PCIe latency efficient GPU caching. 
 To achieve both high performance and user-friendliness, we utilize a blend of C++, CUDA, and Python.
 The data structures of \textbf{tomoCAM} are implemented in C++, while most of the mathematical functions are coded using CUDA. To efficiently handle large datasets, a two-tier partition scheme is employed to seamlessly stream data into and out of GPU memory. To address the vast number of pixels in a typical synchrotron micro- or nano-CT sinogram, which can exceed $\mathcal{O}(10^{10})$, we have carefully optimized the memory usage in the implementation of \textbf{tomoCAM}. For instance, to minimize memory footprint, we pass large arrays that contain frequently accessed data such as the most recent solution, projection data, and gradient as references rather than copies, which is the default behavior in C++.
  We have implemented various strategies to minimize the memory footprint, including:
 \begin{itemize}
     \item Quantities are never stored as complex numbers in the host memory. This additionally helps with the amount of data copied to and from the GPU memory.
     \item Instead of duplicating data, partitions contain pointers to memory locations in the parent array.
     \item Gradients are updated in place when computing the total-variation constraint.
     \item Projection data is reordered into sinogram form for fast contiguous readouts.
 \end{itemize}

 \subsection{GPU Optimizations}
\begin{figure}
    \centering
    \includegraphics[width=0.7\textwidth]{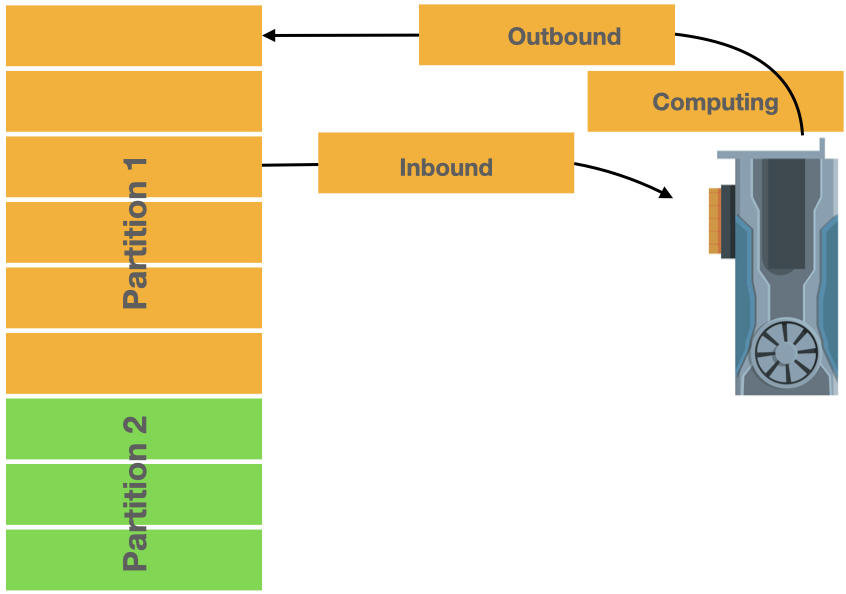}
    \caption{Large arrays are partitioned along the axis of rotation using a two-tier partitioning scheme.}
    \label{fig:async_transfer}
\end{figure}

While GPUs are highly efficient in performing complex calculations, the latency over the PCIe bus remains a significant bottleneck for GPU-accelerated software implementations. In order to minimize runtime and maximize throughput from CPU to GPU memory, we employ a combination of techniques. These include asynchronous transfers, \texttt{OpenMP} threads, and a two-tier data partitioning scheme. The partitioning is done along the axis of rotation, with the data first divided into as many partitions as there are available GPU devices. Each partition is then further subdivided into smaller chunks, with the optimal size depending on the GPU device's available memory. The sub-partitions are streamed to GPU memory, and to minimize memory footprint, they do not create deep copies of the data. Figure \ref{fig:async_transfer} provides an overview of this process. By utilizing these techniques, we can significantly reduce the impact of the PCIe bottleneck and achieve higher performance in our GPU-accelerated software implementations. Some of the other optimizations and features of tomoCAM include:

\begin{itemize}
 \item Since the axis of rotation may not be aligned with the center of the image, we use the Fourier shift property to efficiently move the rotation axis to the center of the image.
\item We use \texttt{OpenMP} threads to parallelly launch level-1 partitions on all the available GPUs, as well as to stream data into GPU memory.
\item  To improve cache efficiency, we utilize GPUs' \texttt{\_\_shared\_\_} memory to store data that is accessed multiple times, such as when computing the \emph{total-variation} constraint.
\item A python front-end and \texttt{numpy} compatibility are provided via pybind11 project \cite{pybind11}. 
\end{itemize}

\textbf{tomoCAM} is publicly available as an open-source project via \url{https://github.com/lbl-camera/tomocam}.

\section{NUMERICAL EXPERIMENTS}
We tested \textbf{tomoCAM} with publicly available phantoms and measured datasets.
Here, we present a comparison of reconstructed results using \textbf{tomoCAM}, 
\textbf{SVMBIR} \cite{svmbir2020} and 
filtered back-projection using \texttt{gridrec} available in the Tomopy package \cite{tomopy}.
Each reconstruction and line-profile ($B$) is 
scaled with $s$ and shifted with $\Delta$, where 
$s, \Delta = \argmin_{s,\Delta} \| A - s\,B + \Delta \|$ to the ground truth ($A$) before plotting. 
In the case of experimental data, we rescale reconstructions from \textbf{SVMBIR} and \texttt{gridrec}
with the one obtained from \textbf{tomoCAM}.
The total-variation constraint used in \textbf{SVMBIR} is slightly different 
from the one used in \textbf{tomoCAM}, see  the theory section in \cite{svmbir2020}. 
\textbf{SVMBIR} uses 10 nearest neighbors,
while the \textbf{tomoCAM} uses 26 of them, to evaluate \eqref{eqn:reg2}. 
We believe parameters can be fine-tuned 
for \textbf{tomoCAM} and \textbf{SVMBIR} to produce equivalent results. 
The primary comparison with \textbf{SVMBIR} is to demonstrate
performance gains, rather than comparing two different constraints or image quality.
All the tests were done on a single machine with
\begin{itemize}
    \item 2 $\times$ Intel(R) Xeon(R) CPU E5-2620 v4 @ 2.10GHz
    \item 4 $\times$ Tesla P100 GPUs
    \item 128 GB RAM
\end{itemize}

\begin{figure}%
\centering%
\begin{minipage}{0.25\linewidth}%
\centering
\includegraphics[width=\textwidth]{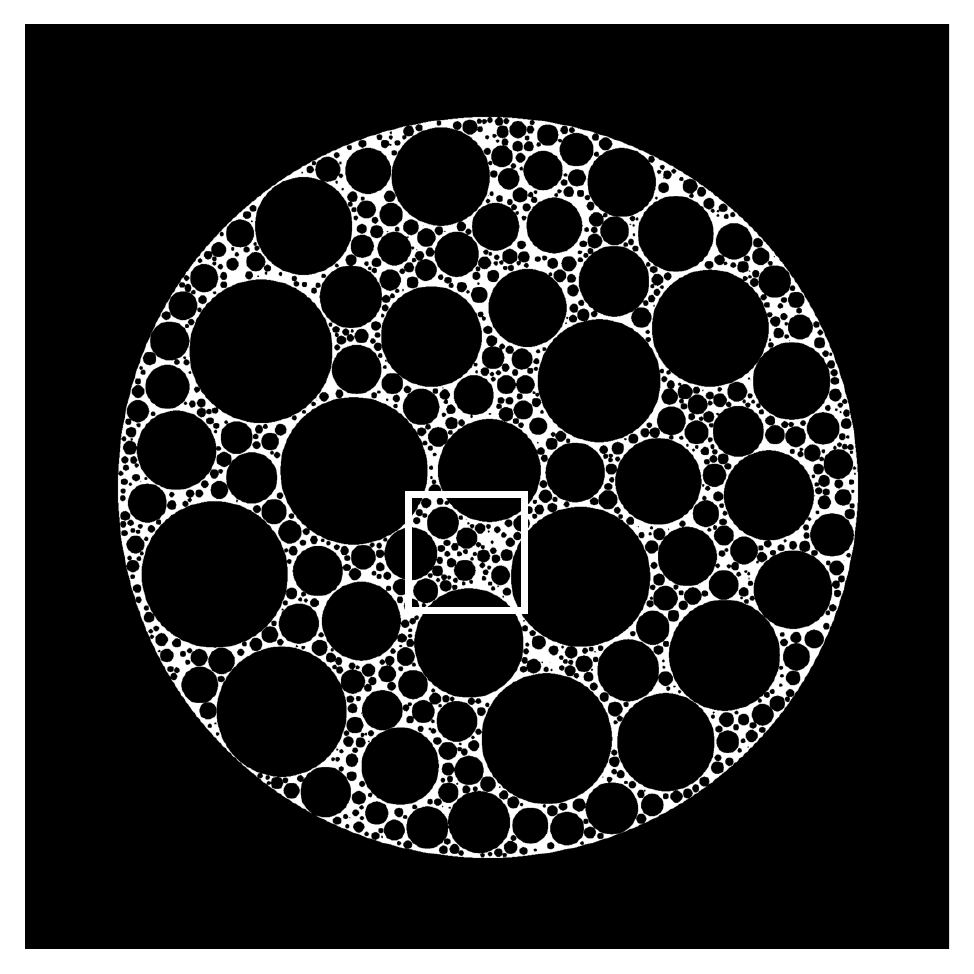}
(a)
\end{minipage}%
\begin{minipage}{0.25\linewidth}%
\centering
\includegraphics[width=\textwidth]{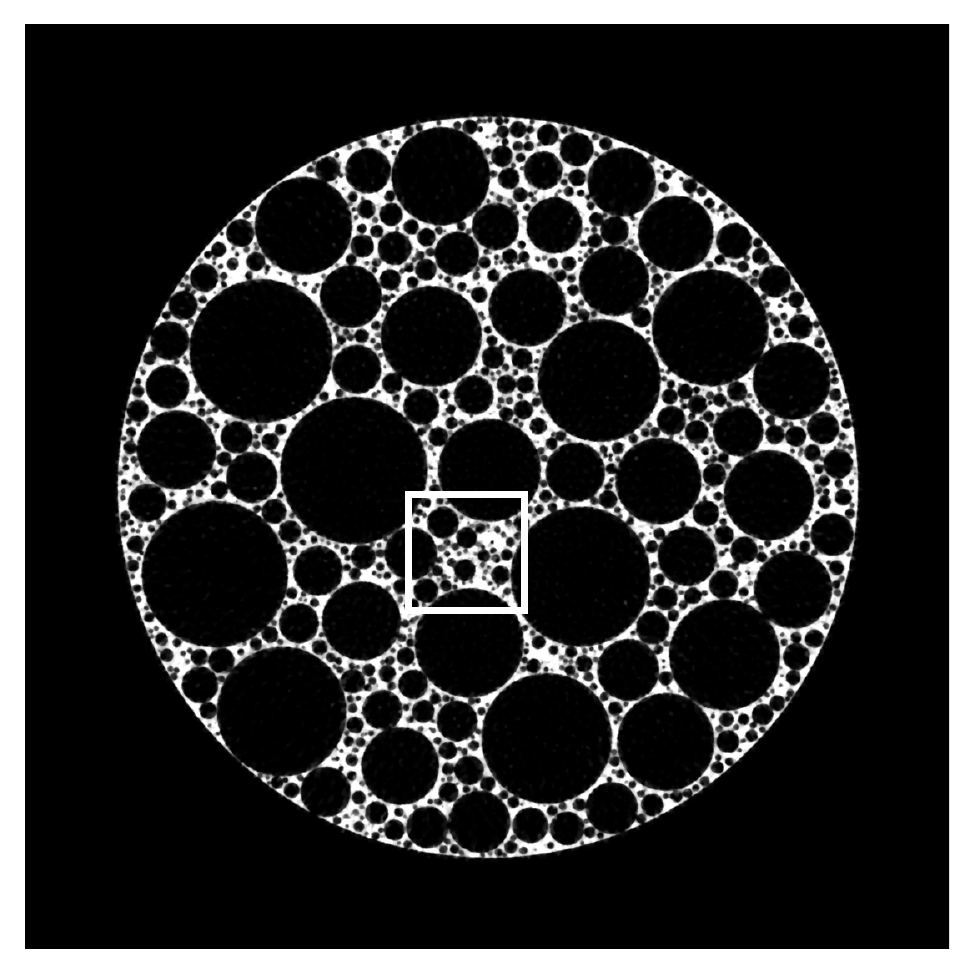}
(b)
\end{minipage}%
\begin{minipage}{0.25\linewidth}%
\centering
\includegraphics[width=\textwidth]{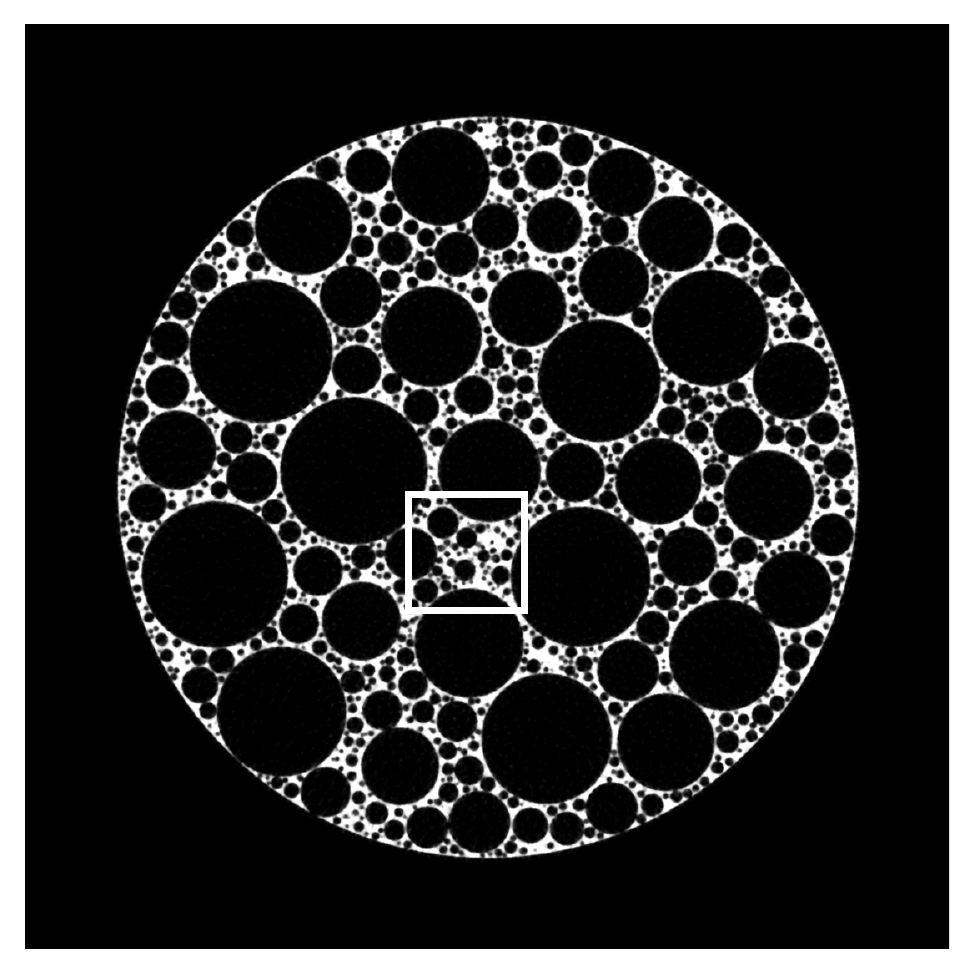}
(c)
\end{minipage}%
\begin{minipage}{0.25\linewidth}%
\centering
\includegraphics[width=\textwidth]{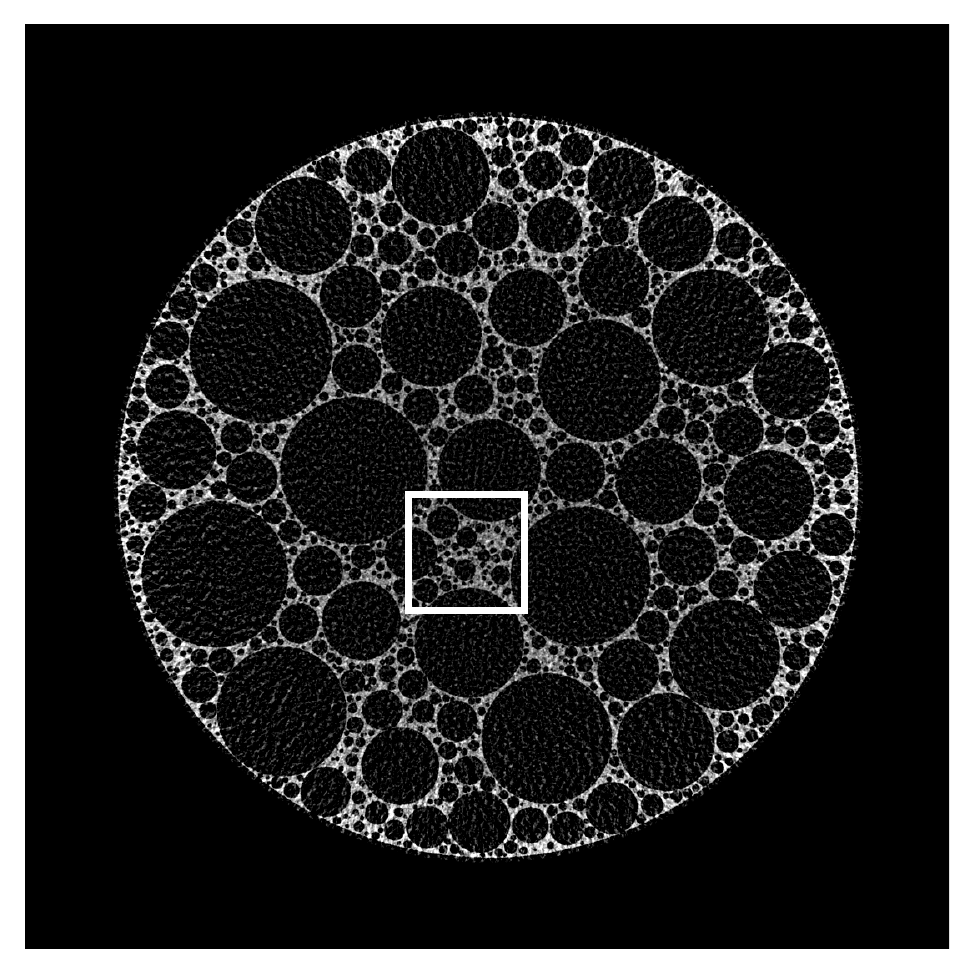}
(d)
\end{minipage}%

\begin{minipage}{0.25\linewidth}%
\centering
\includegraphics[width=\textwidth]{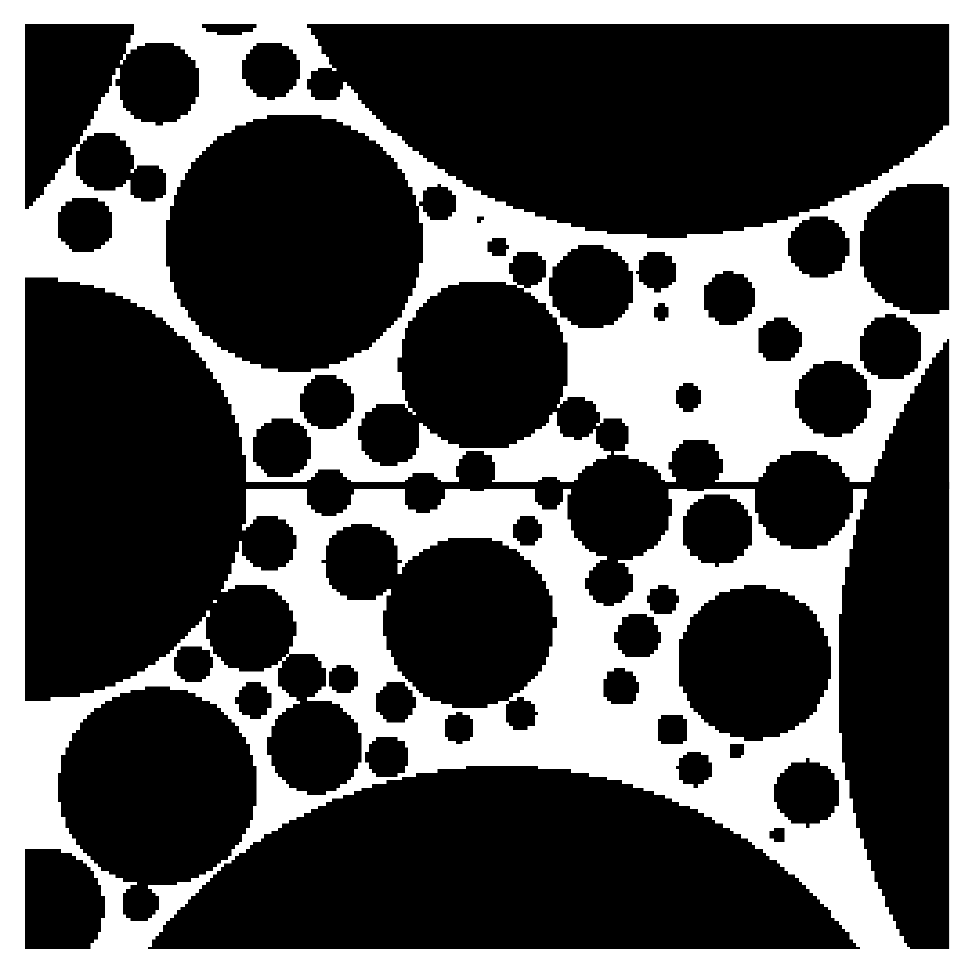}
(e)
\end{minipage}%
\begin{minipage}{0.25\linewidth}%
\centering
\includegraphics[width=\linewidth]{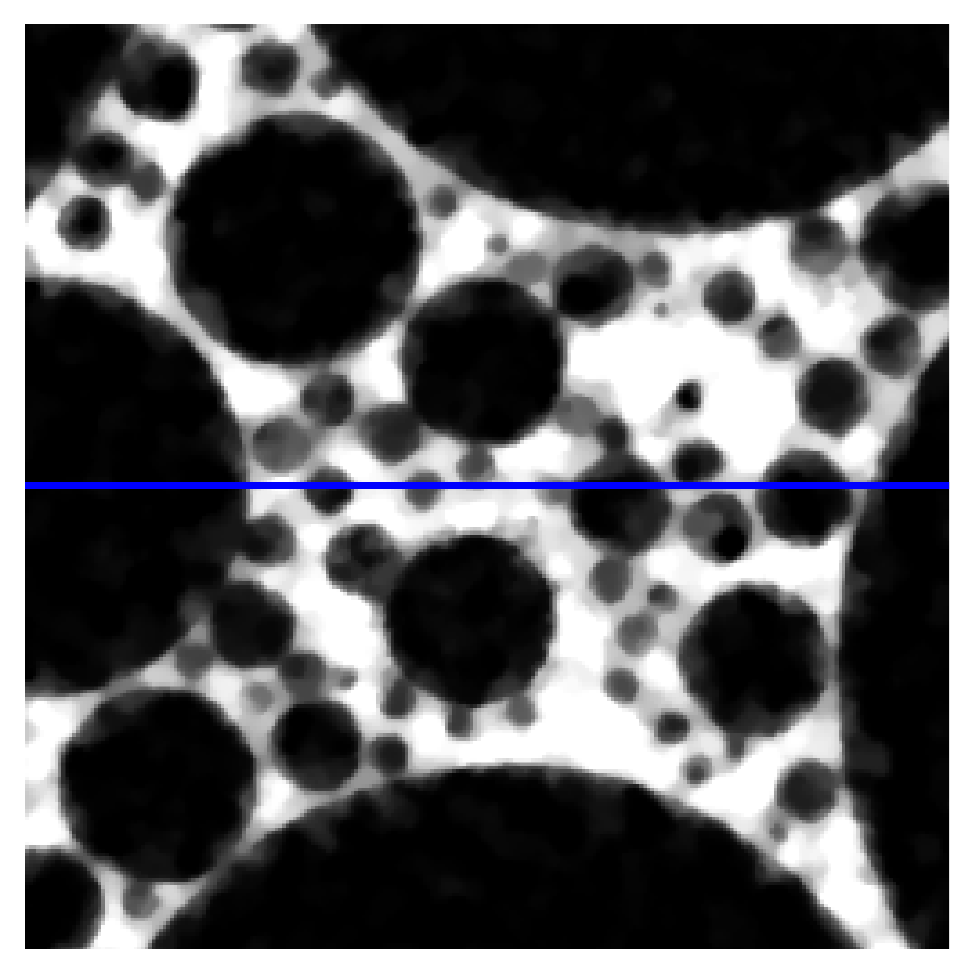}
(f)
\end{minipage}%
\begin{minipage}{0.25\linewidth}%
\centering
\includegraphics[width=\linewidth]{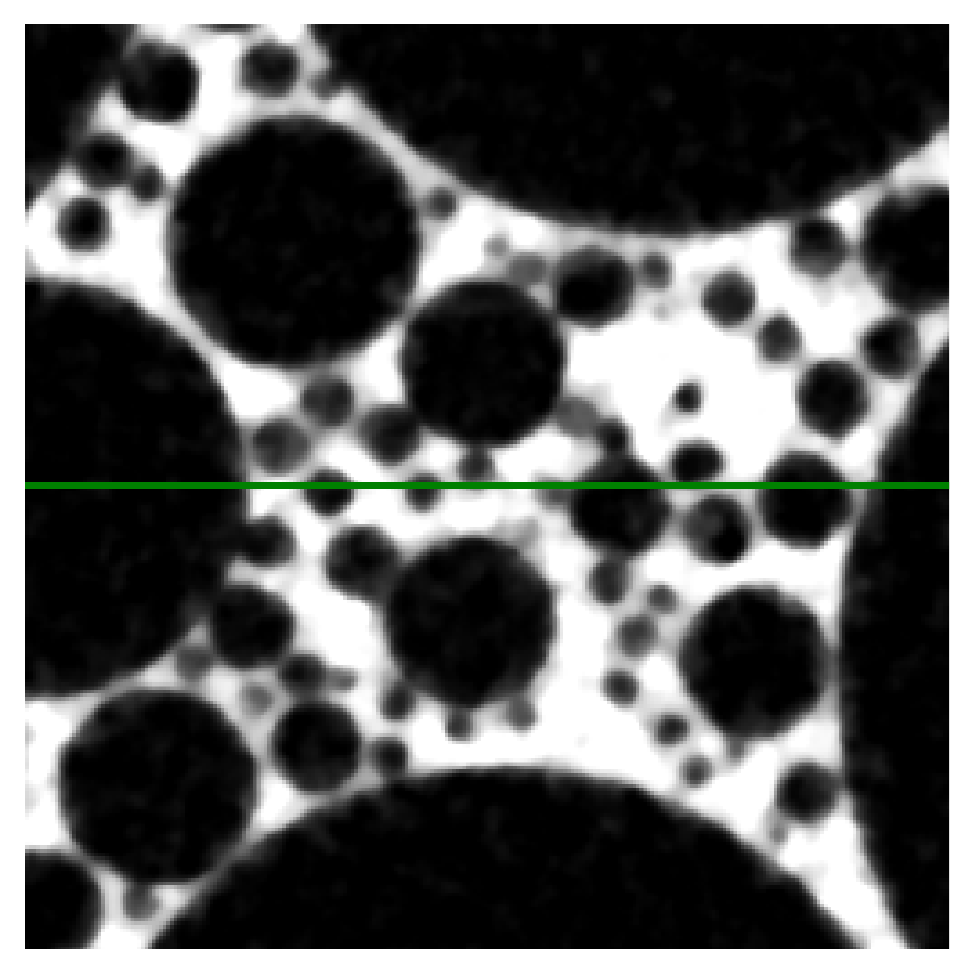}
(g)
\end{minipage}%
\begin{minipage}{0.25\linewidth}%
\centering
\includegraphics[width=\linewidth]{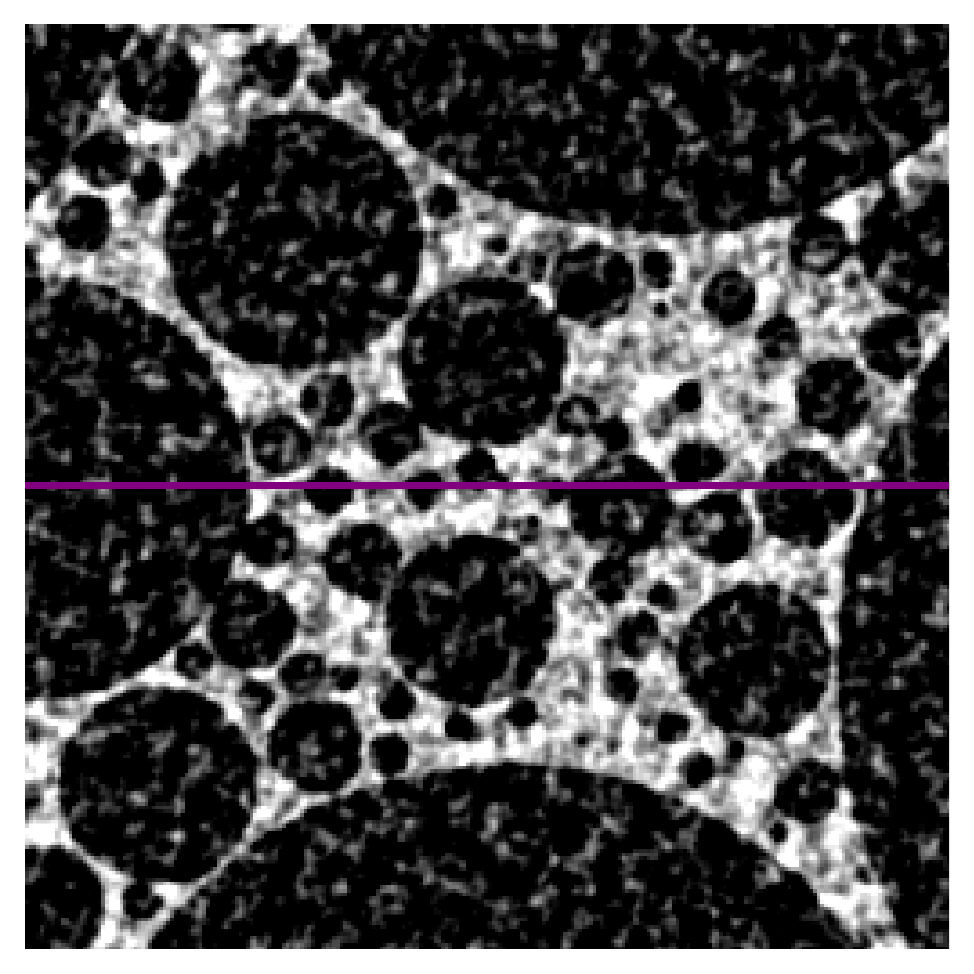}
(h)
\end{minipage}\\
\begin{minipage}{\linewidth}%
\centering
\includegraphics[width=\linewidth]{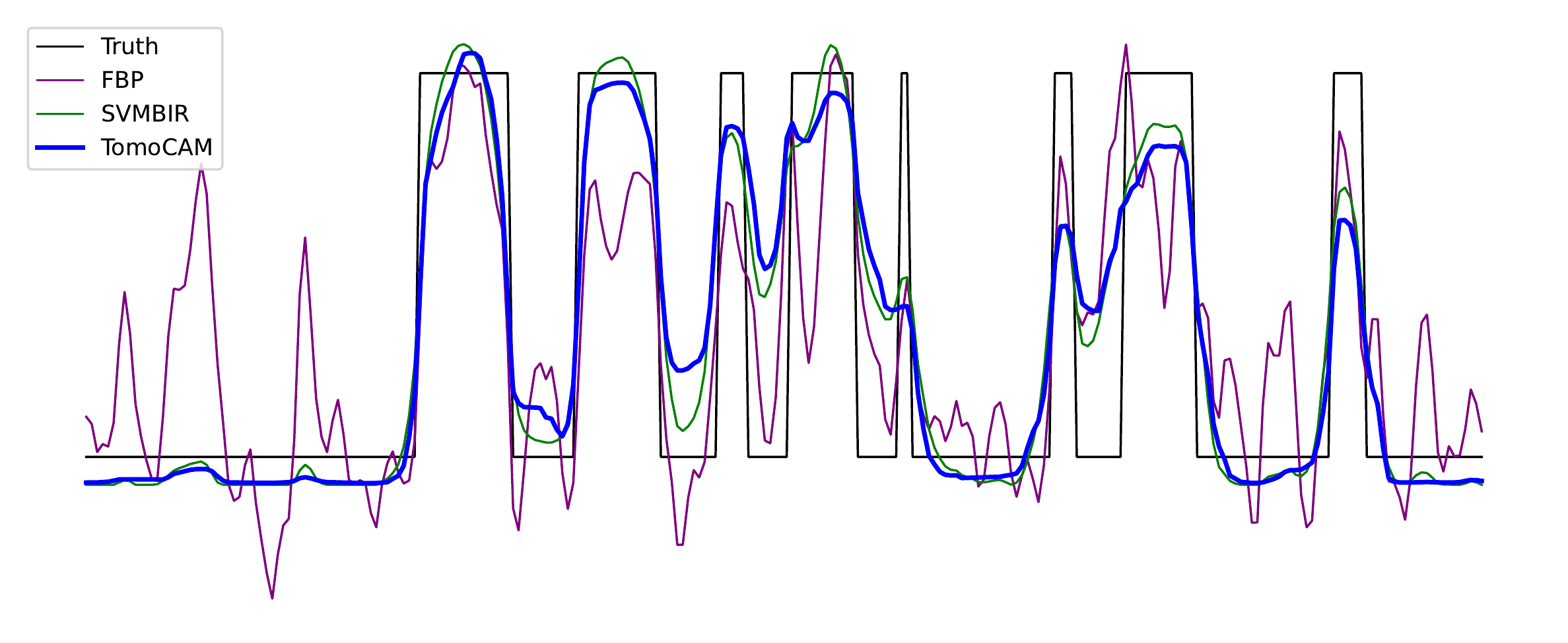}
(i)
\end{minipage}

\caption{A compassion of reconstruction methods for a foam phantom with 128 projections
(a) Ground Truth,
(b) \textbf{tomoCAM},
(c) \textbf{SVMBIR}, and
(d) \texttt{gridrec}.
(e), (f), (g) and (h) are the zoomed-in regions of interest represented by the boxes in (a), (b), (d) and (e) respectively, and 
(i) displays the line profiles on (e), (f), (g) and (h). Reconstructions using \textbf{tomoCAM} and \textbf{SVMBIR} 
result in images with low noise, when compared to \texttt{gridrec}. Here \textbf{tomoCAM} is about 9$\times$ faster than \textbf{SVMBIR}.}
\label{fig:foam_phantom}
\end{figure}

In the first experiment, we compare the reconstruction of a foam phantom from all three codes.
A foam phantom and its projection data of size $(128 \times 16 \times 2048) $ was generated  
using the \textbf{foam\_ct\_phantom} package \cite{Pelt2022}. 
A full slice from the phantom in fig  \ref{fig:foam_phantom}(a), is compared with
the reconstruction obtained from each code (\textbf{SVMBIR}, \textbf{tomoCAM}, and \texttt{gridrec}) in \ref{fig:foam_phantom}(b-d). This is followed by \emph{zoomed-in} regions of each image in \ref{fig:foam_phantom}(e-h). A line profile from each of the 
zoomed-in regions is then compared in \ref{fig:foam_phantom}(i). 
It is evident from the results, that both \textbf{tomoCAM} and \textbf{SVMBIR} 
are effective at suppressing the noise. One major advantage of \textbf{tomoCAM} is that it can get equivalent results in an order of magnitude faster time.

Next, we evaluate the reconstruction of two experimental datasets obtained from diverse synchrotron light-sources that are accessible through Tomobank \cite{tomobank2018}. 
For each dataset, the available number of projections is notably lower than what is typically expected, which follows the general rule of thumb that it should be as many as the number of pixel columns in the camera sensor. 

Using \emph{Beer-Lambert's law} \cite{Swinehart1962} $b$ in equation \eqref{eqn:objfn} is defined as  
$- \log(I/I_0)$, where $I$ is the measured intensity and
$I_0$ is the beam intensity without the sample blocking the view.
 The hyper parameters used for \textbf{tomoCAM} are, $p=1.2$, $\sigma=\frac{1}{0.7}$ and $c = 0.0001$. 
 For the \texttt{gridrec} we chose the Butterworth filter with order $2$, and the cutoff frequency was set to $0.25$, which is typical for a synchrotron tomographic reconstruction.  We choose $T=1$ and 
 $\sigma_x = \{0.98,\, 2.1,\, 1.1\} \times 10^{-4}$ for the phantom, Tomobank dataset id 25 (TB-25), and Tomobank dataset id 86 (TB-86) respectively for \textbf{SVMBIR} runs, in order to produce similar quality reconstructions as \textbf{tomoCAM}.

 We follow a similar pattern to Fig.\@ \ref{fig:foam_phantom} for plotting images
and line profiles. The first row of images depicts full slices, followed by zoomed-in regions, and then
a line profile is taken from the middle of each zoomed-in region. We expect that conducting a thorough hyper-parameter search would yield comparable outcomes from both \textbf{tomoCAM} and \textbf{SVMBIR}, given their mathematical similarity. Table \ref{tab:runtimes} shows a comparison of the time taken by each code.

\begin{table}[]
    \centering
    \begin{tabular}{lcccc}
    \hline
    & \multicolumn{3}{c}{Reconstruction Time (s)} \\
        & & \textbf{tomoCAM} & \textbf{SVMBIR} & \texttt{gridrec} \\
    Dataset & size &&& \\
     \hline
    Phantom & $(128, 16,2048)$ & 93 & 810 & 0.21 \\
    TB-25 & $(400, 128, 2048)$ & 862 & 12730 & 2.53 \\
    TB-86 & $(202, 128, 2448)$ & 1210 & 14273 & 3.73 \\
    \hline \\
    \end{tabular}\\
    \vspace{2pt}
    \caption{A comparison of time taken to reconstruct various datasets. Both \textbf{tomoCAM} and \textbf{SVMBIR} were timed for 100 iterations.}
    \label{tab:runtimes}
\end{table}

\begin{figure}%
\centering%
\begin{minipage}{0.33\linewidth}%
\centering
\includegraphics[width=\textwidth]{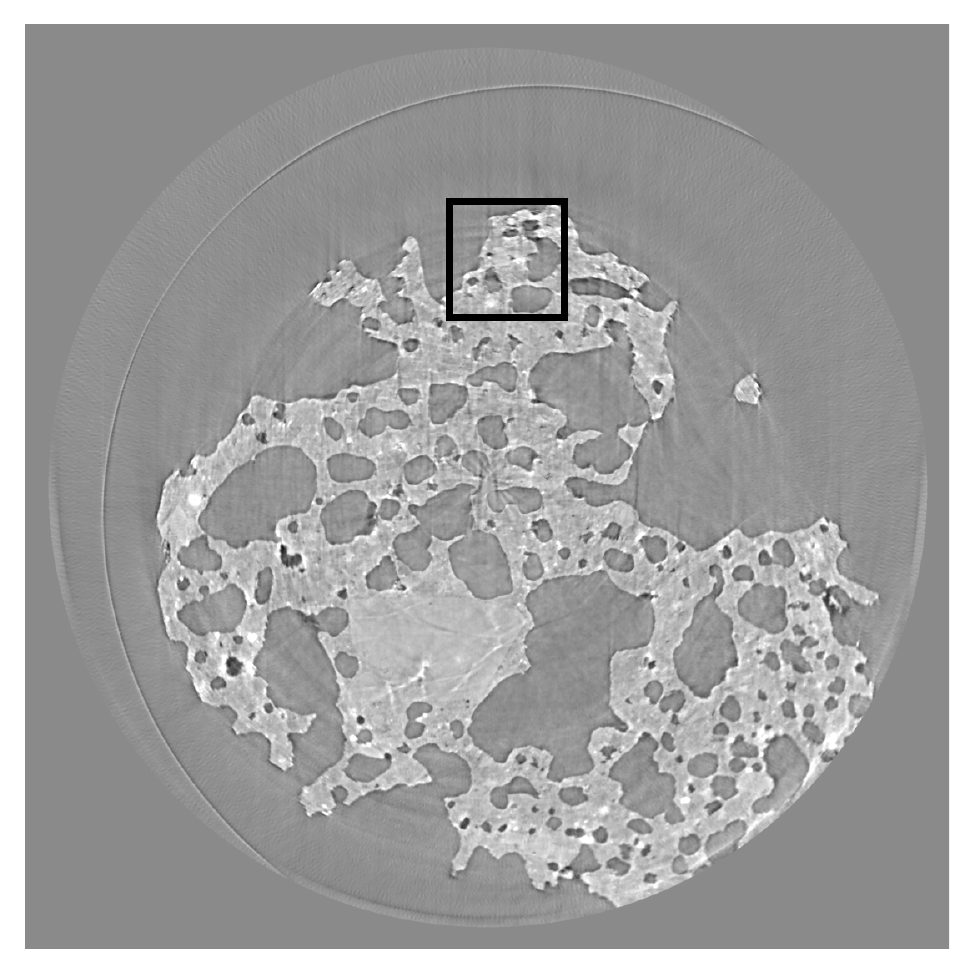}
(a)
\end{minipage}%
\begin{minipage}{0.33\linewidth}%
\centering
\includegraphics[width=\textwidth]{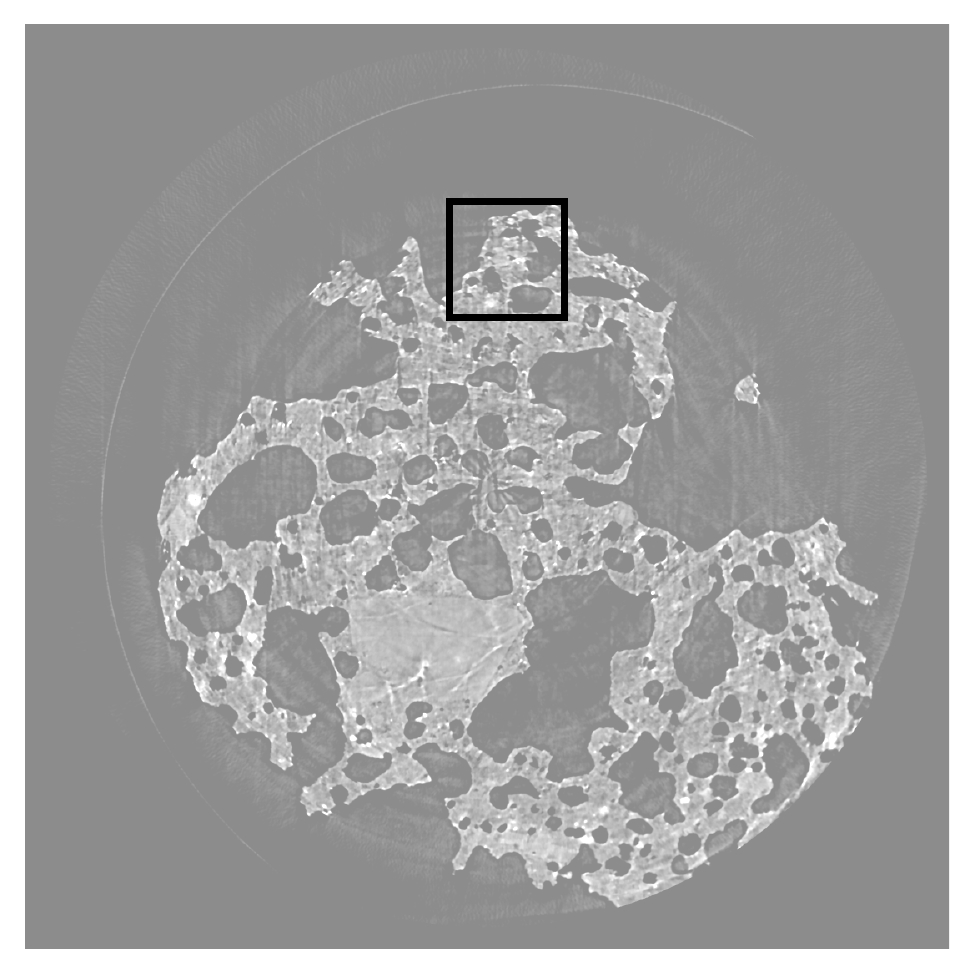}
(b)
\end{minipage}%
\begin{minipage}{0.33\linewidth}%
\centering
\includegraphics[width=\textwidth]{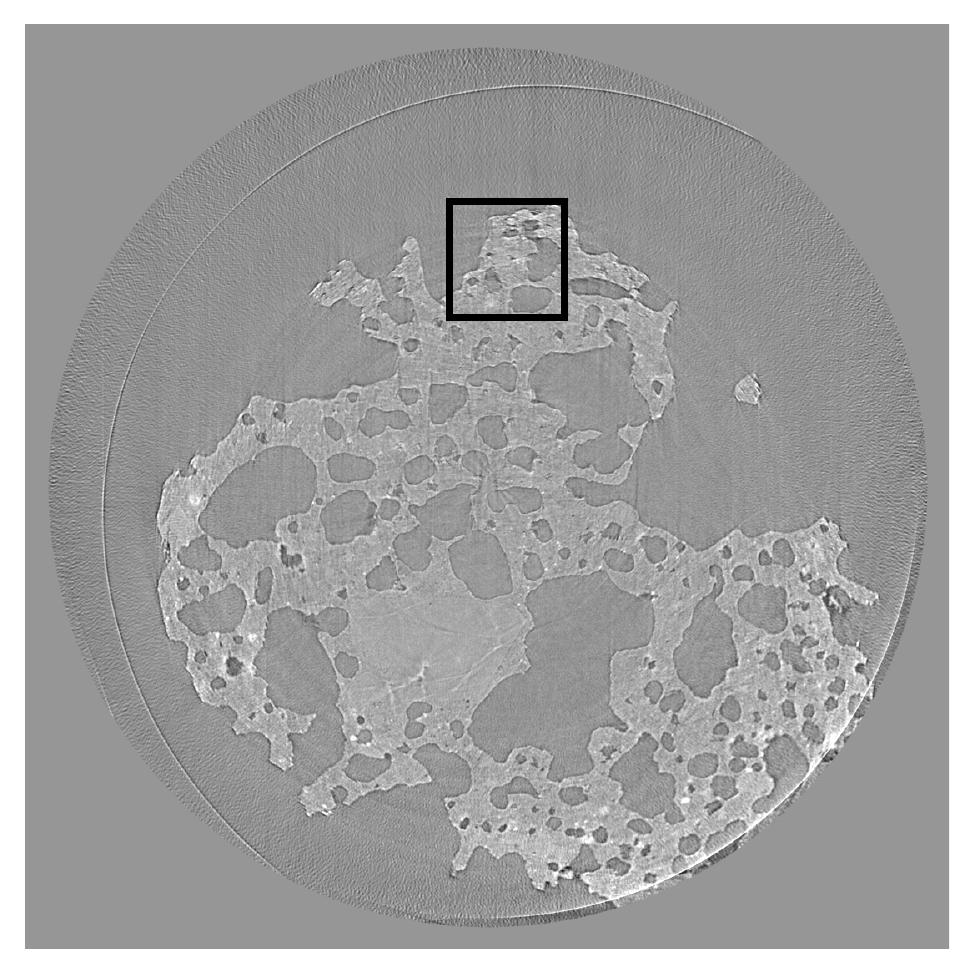}
(c)
\end{minipage}\\
\begin{minipage}{0.33\linewidth}%
\centering
\includegraphics[width=\textwidth]{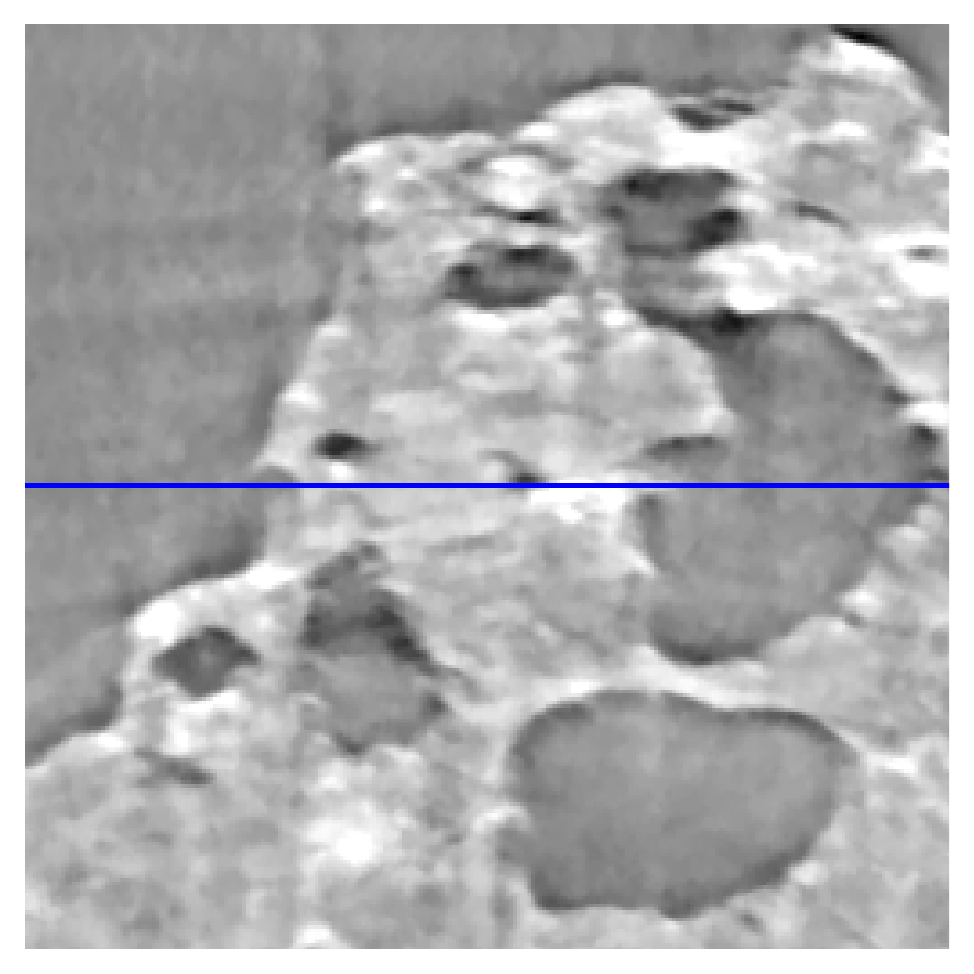}
(d)
\end{minipage}%
\begin{minipage}{0.33\linewidth}%
\centering
\includegraphics[width=\textwidth]{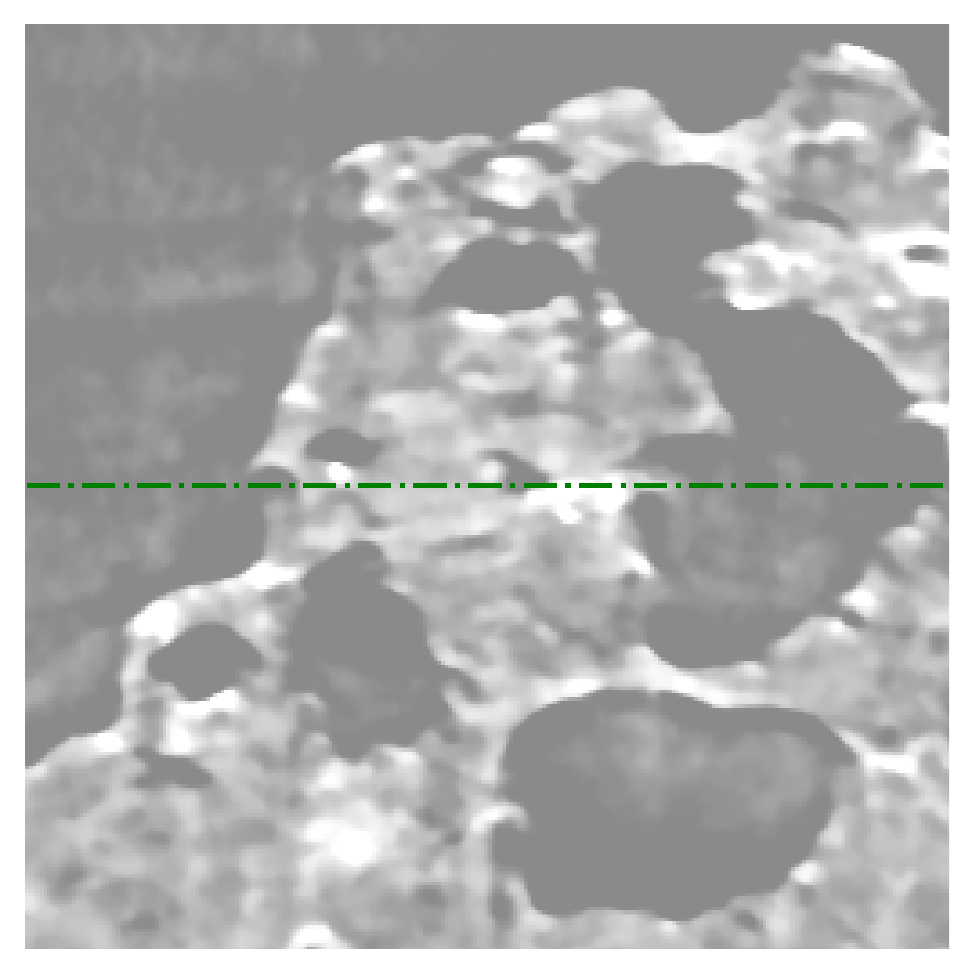}
(e)
\end{minipage}%
\begin{minipage}{0.33\linewidth}%
\centering
\includegraphics[width=\textwidth]{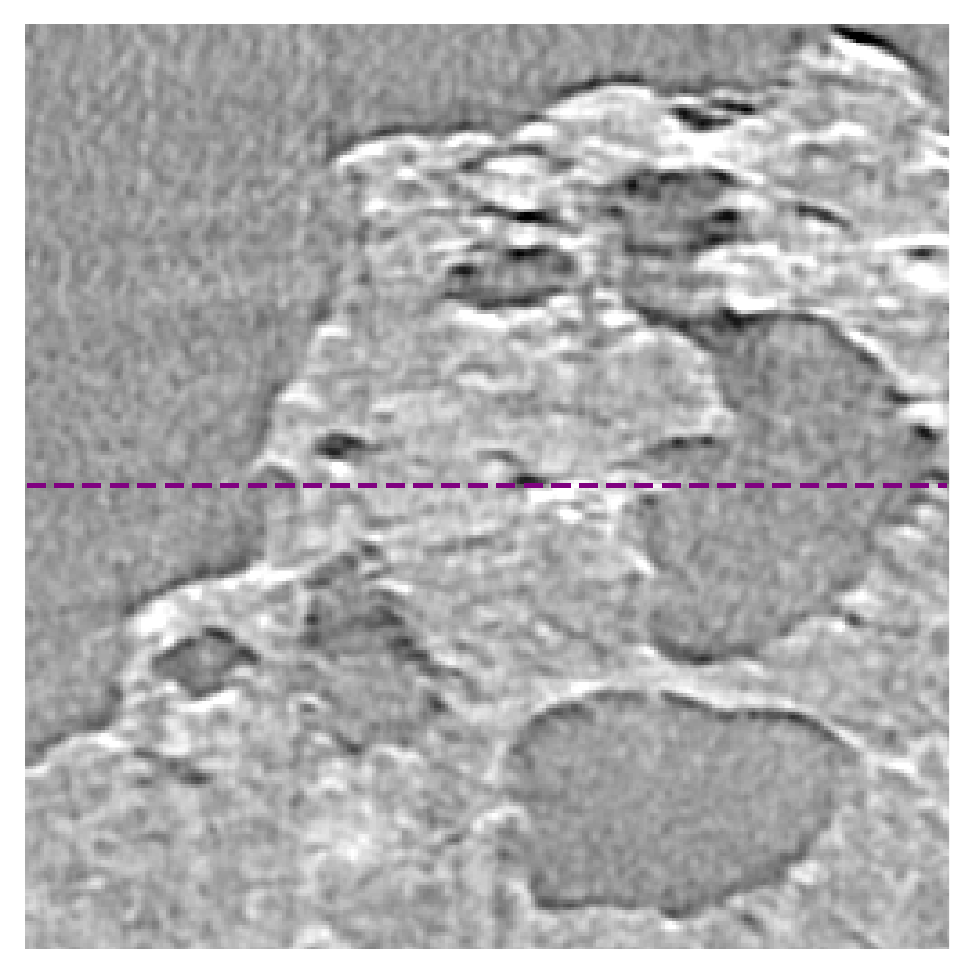}
(f)
\end{minipage}\\
\begin{minipage}{\linewidth}
\centering
\includegraphics[width=\textwidth]{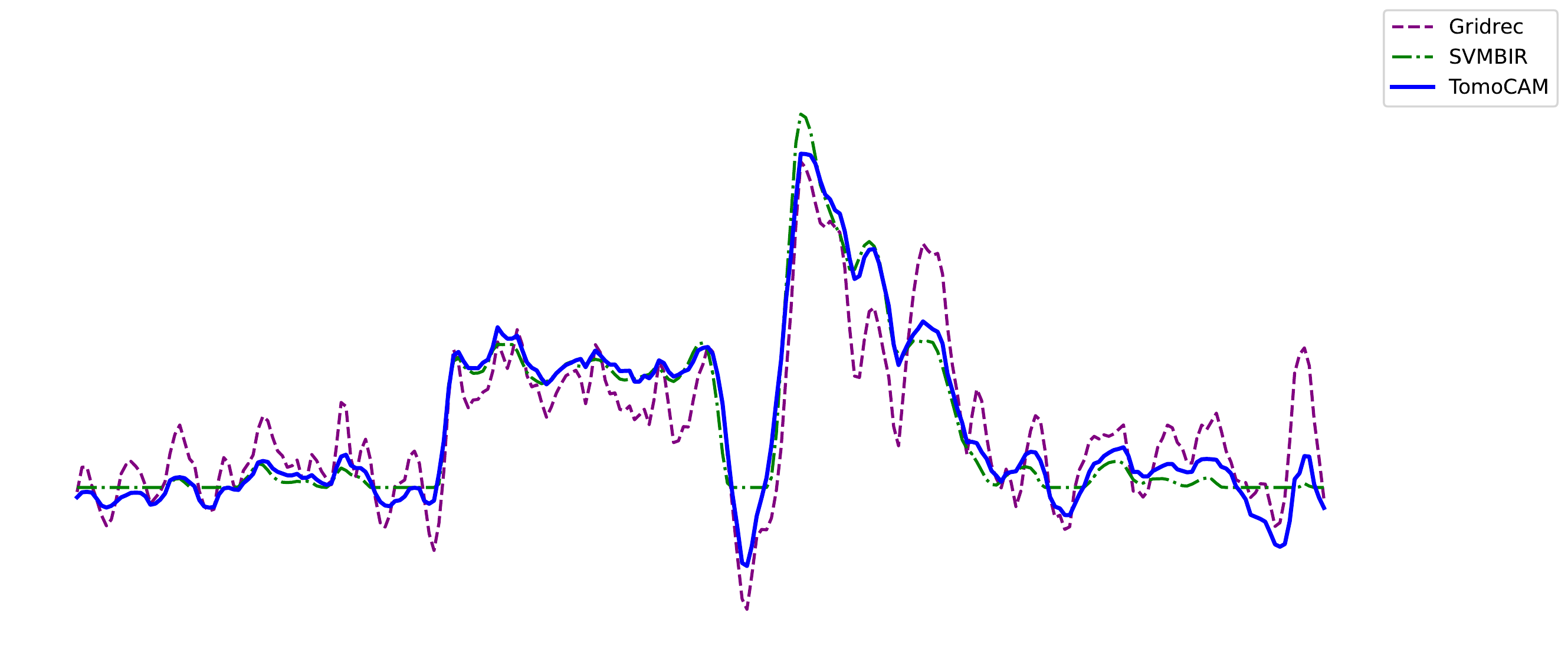}
(g)
\end{minipage}

\caption{Reconstructions for Tomobank Dataset ID: 25, \emph{in-situ} study of rock permeability with 400 projections, with
(a) \textbf{TomoCAM}, 
(b) \textbf{SVMBIR}, and
(c) \texttt{gridrec}.
(d), (e), and (f) are zoomed-in regions of interest represented by the boxes in (a), (b), and (c) respectively.
(e) displays the line profiles for (d), (e), and (f). 
A circular mask was applied to all the reconstructions. While \textbf{tomoCAM} and \textbf{SVMBIR} both do
an excellent job at suppressing the noise when compared to \texttt{gridrec}, \textbf{tomoCAM} is
approximately 15$\times$ faster than \textbf{SVMBIR}.
}
\label{fig:tb25}
\end{figure}
\begin{figure}%
\centering%
\begin{minipage}{0.33\linewidth}%
\centering
\includegraphics[width=\textwidth]{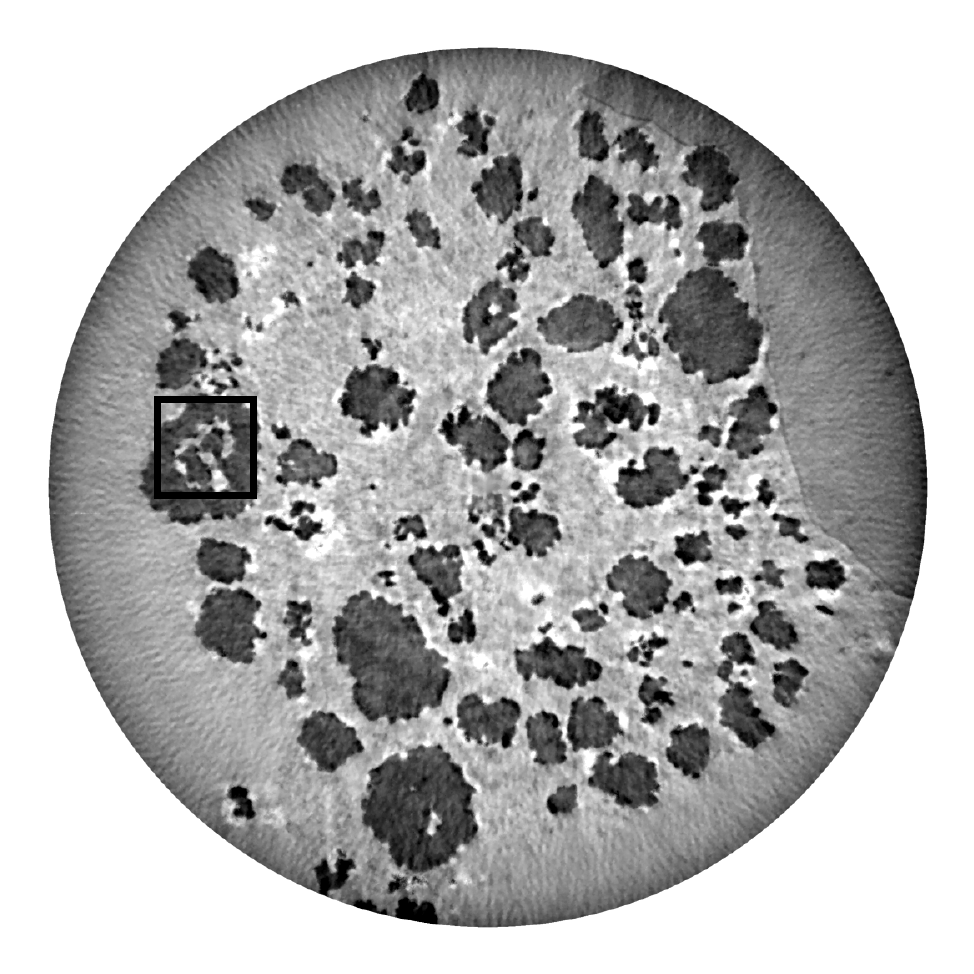}
(a)
\end{minipage}%
\begin{minipage}{0.33\linewidth}%
\centering
\includegraphics[width=\textwidth]{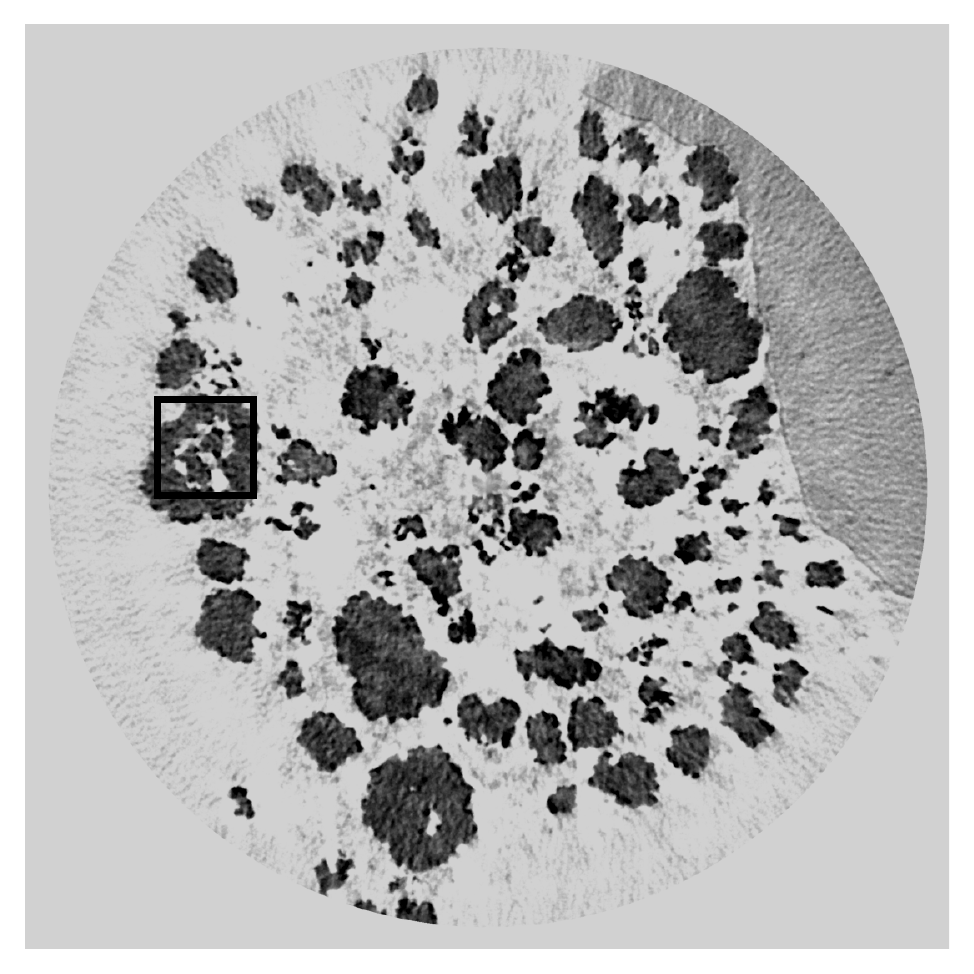}
(b)
\end{minipage}%
\begin{minipage}{0.33\linewidth}%
\centering
\includegraphics[width=\textwidth]{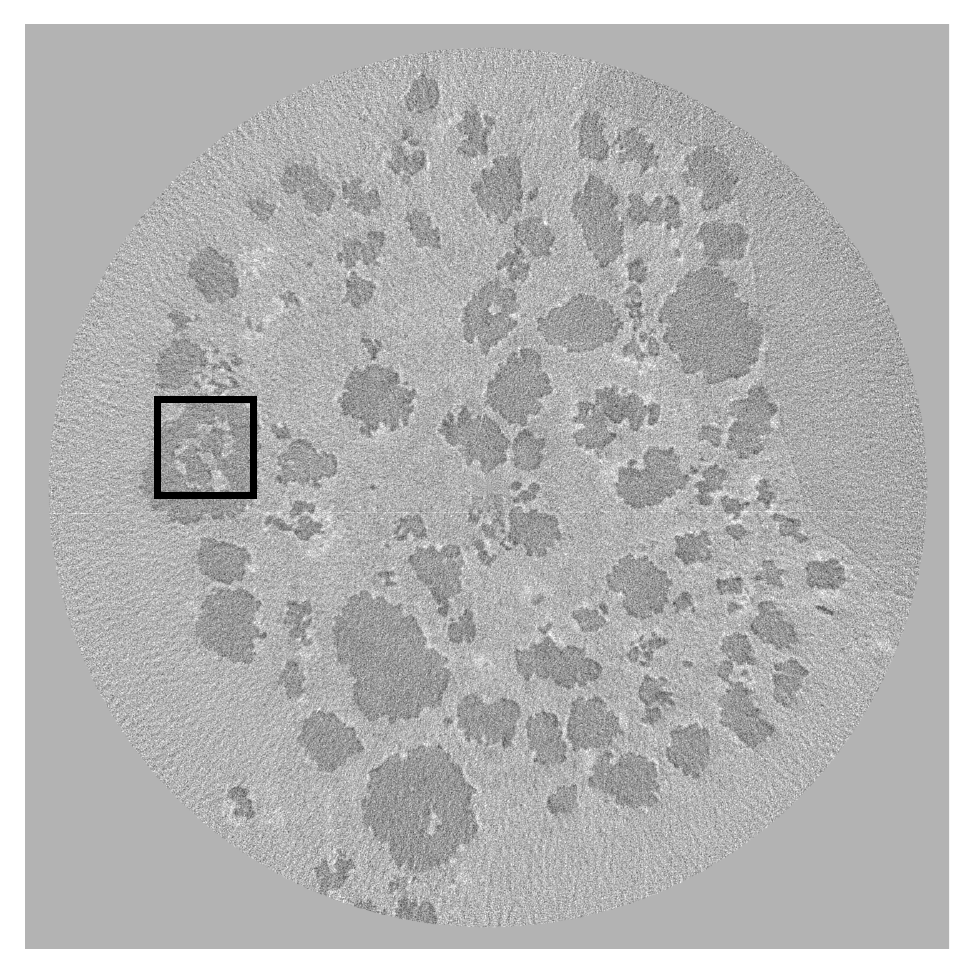}
(c)
\end{minipage}\\
\begin{minipage}{0.33\linewidth}%
\centering
\includegraphics[width=\textwidth]{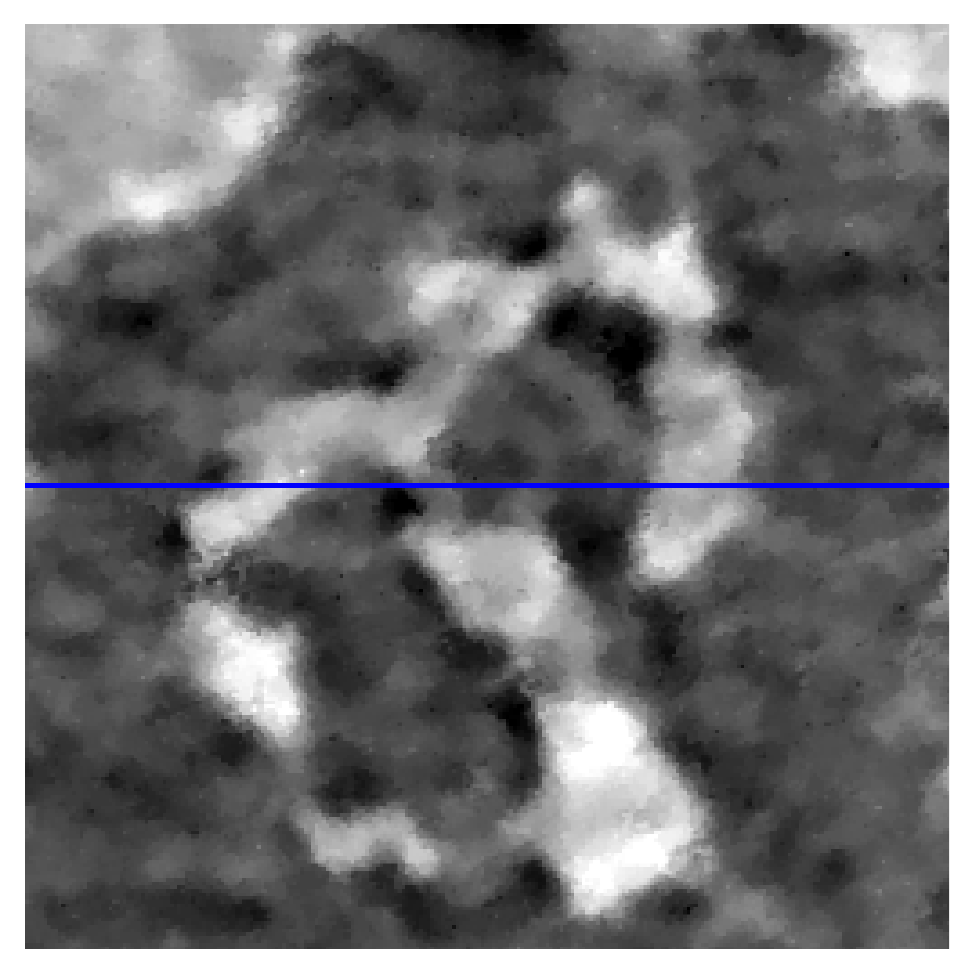}
(d)
\end{minipage}%
\begin{minipage}{0.33\linewidth}%
\centering
\includegraphics[width=\textwidth]{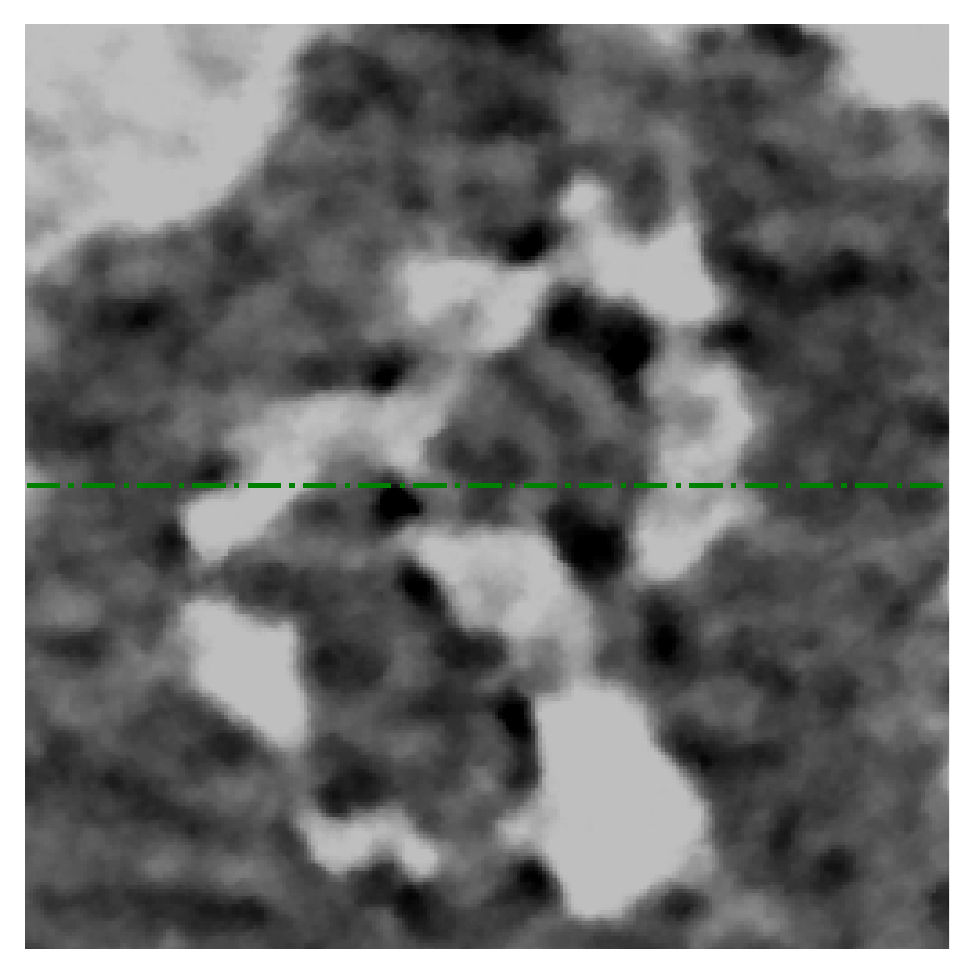}
(e)
\end{minipage}%
\begin{minipage}{0.33\linewidth}%
\centering
\includegraphics[width=\textwidth]{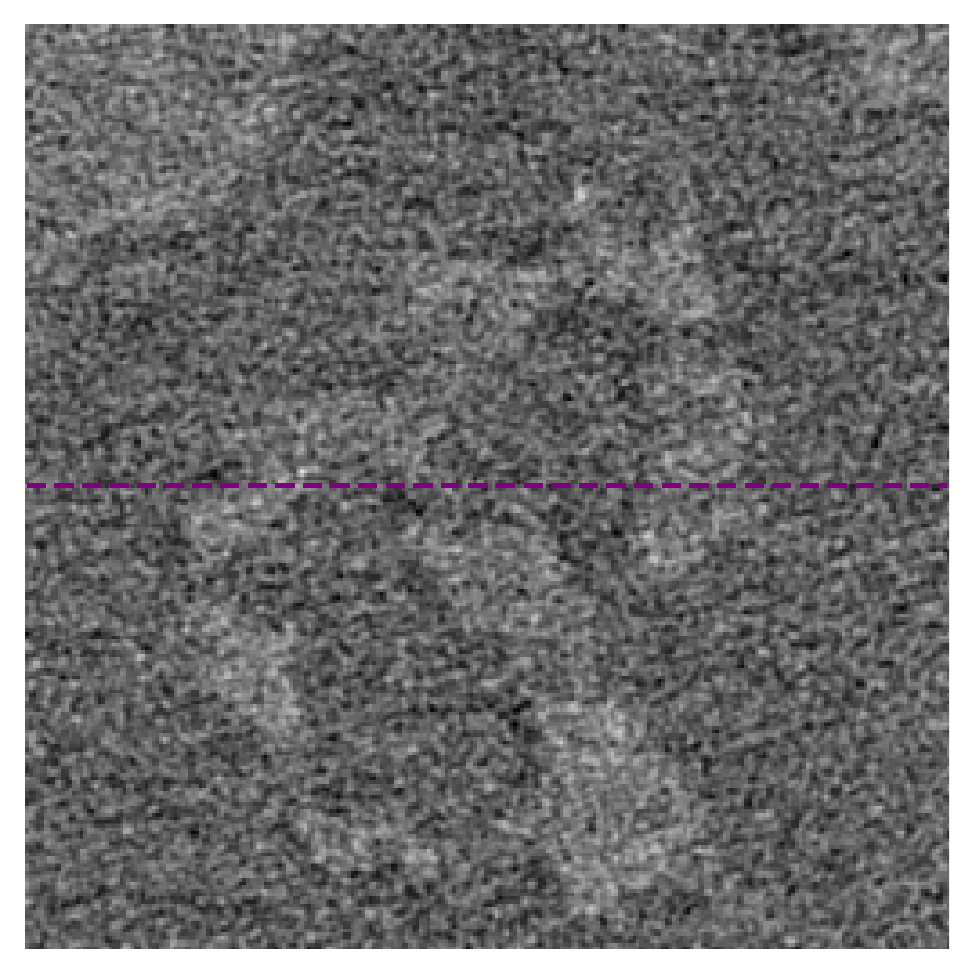}
(f)
\end{minipage}\\
\begin{minipage}{\linewidth}
\centering
\includegraphics[width=\textwidth]{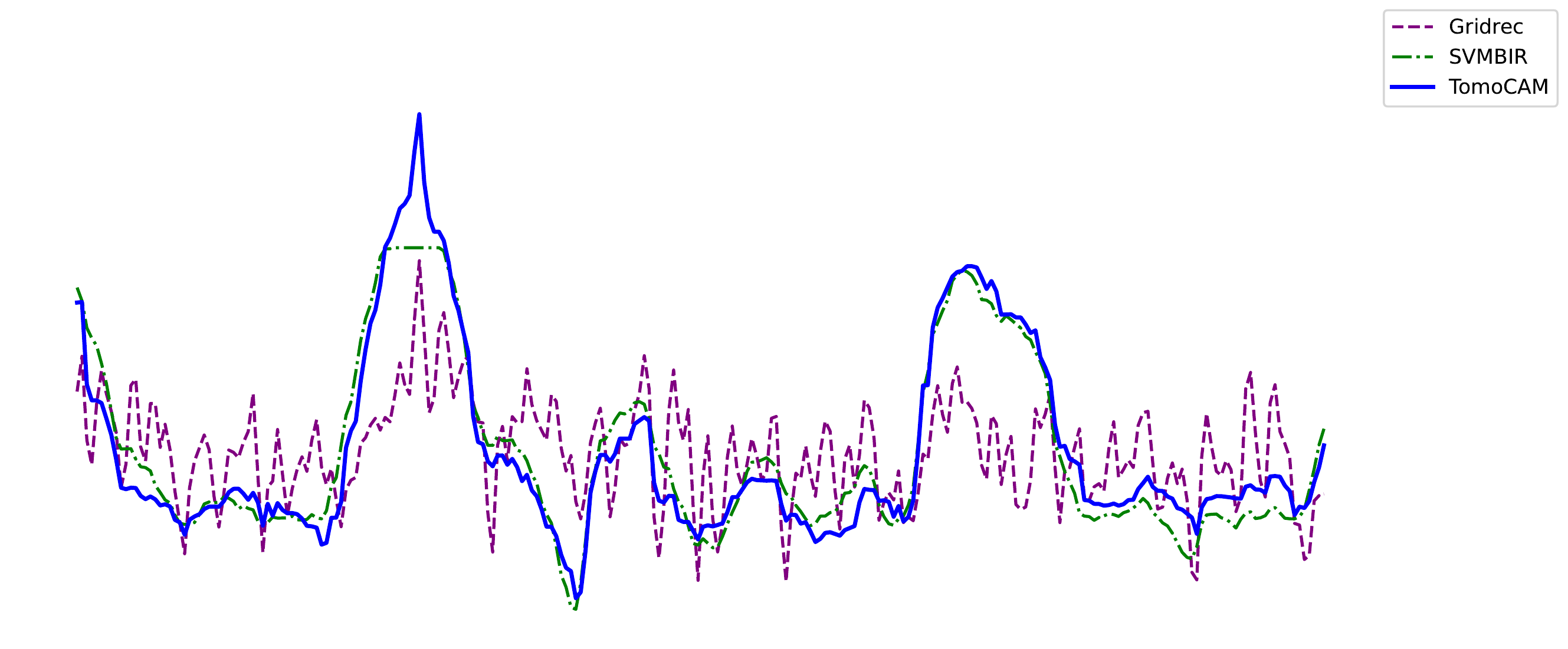}
(g)
\end{minipage}

\caption{Reconstructions for Tomobank Dataset ID: 86, Nano-CT data with sparse projection angles using 202 projections, with
(a) \textbf{tomoCAM}, 
(b) \textbf{SVMBIR}, and
(c) \texttt{gridrec}.
(d), (e), and (f) are zoomed-in regions of interest represented by the boxes in (a), (b), and (c) respectively.
(e) displays the line profiles for (d), (e), and (f). 
A circular mas was applied to all the reconstructions.
 While \textbf{tomoCAM} and \textbf{SVMBIR} both do
an excellent job at suppressing the noise when compared to \texttt{gridrec}, \textbf{tomoCAM} is
approximately 15$\times$ faster.
}
\label{fig:tb86}
\end{figure}

\section{CONCLUSIONS}
In this work, we have presented tomoCAM, a new GPU-accelerated software for reconstructing high-quality
tomographic images. \textbf{tomoCAM} is capable of running model-based iterative reconstructions for 
large datasets with relatively modest hardware requirements, within a reasonable time.
The resulting reconstructed images have lower noise when compared with the prevalent  
filtered-back projection methods, while being an order of magnitude faster than CPU-only MBIR implementations. 

A Python-based front-end has been created for \textbf{tomoCAM}, which is specifically designed to receive Numpy arrays as both input and output for reconstructions. This facilitates seamless integration of \textbf{tomoCAM} into the existing workflows of beamline scientists. Although the use of MBIR is particularly advantageous in cases where there is a scarcity of available projection data, the current implementation of MBIR is quite time-consuming. Consequently, this is the primary reason why beamline scientists do not utilize MBIR even when it is obviously advantageous. \textbf{tomoCAM} overcomes this problem, thus making MBIR reconstruction more practical, by

\begin{itemize}
    \item improving efficiency: the run time has been reduced by an order of magnitude, making it faster than previous MBIR versions,
    \item reducing hardware requirements: it can run on machines as small as an individual desktop with a GPU, making it more accessible,
    \item simplifying hyper-parameter search: \textbf{tomoCAM}'s speed makes it easier to search for hyper-parameters, allowing for faster and more efficient experimentation,
    \item enhancing compatibility: the implementation provides a Python interface, which makes it easy to integrate with existing workflows that use FBP.
\end{itemize}


\vspace{5pt}
\ack{\textbf{ACKNOWLEDGEMENTS}}
This work was supported by the Center for Advanced Mathematics for Energy Research Applications (CAMERA), funded by the Advanced Scientific
Computing Research and Basic Energy Sciences programs
of the Office of Science of the Department of Energy (DOE)
(Award No. DE-AC02-05CH11231).
 We thank S.V. Venkatakrishnan of Oak Ridge National Laboratory for numerous 
 valuable discussions. We would also like to thank our colleagues J.A. Sethian, Z. Hu, and K. Pande,
 for reviewing the manuscript.

\referencelist[main]

\end{document}